\documentclass[iop]{emulateapj}
\usepackage{amsmath}
\usepackage{graphicx}
\usepackage{hyperref}
\usepackage{IEEEtrantools}
\usepackage{mathtools}

\newcommand{\amax}{\ensuremath{A_{\rm max}}}

\newcommand{\deltaF}{\ensuremath{\Delta F}}

\newcommand{\dl}{\ensuremath{D_{\ell}}}

\newcommand{\dproj}{\ensuremath{D_{\perp}}}
\newcommand{\dprojvec}{\ensuremath{\boldsymbol{D}_{\perp}}}
\newcommand{\ds}{\ensuremath{D_{s}}}
\newcommand{\fb}{\ensuremath{F_{b}}}
\newcommand{\fl}{\ensuremath{F_{\ell}}}
\newcommand{\fsource}{\ensuremath{F_{s}}}

\newcommand{\ftbase}{\ensuremath{F_{t, {\rm base}}}}

\newcommand{\hbhighres}{\ensuremath{H_{b, \rm{high-res}}}}
\newcommand{\hl}{\ensuremath{H_{l}}}
\newcommand{\hs}{\ensuremath{H_{s}}}
\newcommand{\hshighres}{\ensuremath{H_{s, {\rm high-res}}}}

\newcommand{\htar}{\ensuremath{H_{t}}}
\newcommand{\htbase}{\ensuremath{H_{t, {\rm base}}}}
\newcommand{\htmag}{\ensuremath{H_{t, {\rm mag}}}}

\newcommand{\mearth}{\ensuremath{{\rm M}_{\oplus}}}
\newcommand{\mffp}{\ensuremath{M_{\rm FFP}}}
\newcommand{\mjup}{\ensuremath{{\rm M}_{\rm Jup}}}
\newcommand{\ml}{\ensuremath{M_{\ell}}}

\newcommand{\msun}{\ensuremath{{\rm M}_{\odot}}}

\newcommand{\murel}{\ensuremath{\mu_{\rm rel}}}
\newcommand{\murelvec}{\ensuremath{\boldsymbol{\mu}_{\rm rel}}}
\newcommand{\piE}{\ensuremath{\pi_{\rm E}}}

\newcommand{\pirel}{\ensuremath{\pi_{\rm rel}}}
\newcommand{\rE}{\ensuremath{r_{\rm E}}}
\newcommand{\rEproj}{\ensuremath{\tilde{r}_{\rm E}}}

\newcommand{\rs}{\ensuremath{R_{s}}}
\newcommand{\rsun}{\ensuremath{{\rm R}_{\odot}}}
\newcommand{\tE}{\ensuremath{t_{\rm E}}}
\newcommand{\thetaE}{\ensuremath{\theta_{\rm E}}}
\newcommand{\thetastar}{\ensuremath{\theta_{*}}}
\newcommand{\tzero}{\ensuremath{t_{\rm 0}}}
\newcommand{\uzero}{\ensuremath{u_{\rm 0}}}

\newcommand{\vl}{\ensuremath{v_{\ell}}}
\newcommand{\vs}{\ensuremath{v_{s}}}
\newcommand{\vsun}{\ensuremath{v_{\odot}}}

\newcommand{\euclid}{\it Euclid}
\newcommand{\gaia}{\it Gaia}
\newcommand{\newhorizons}{\it New Horizons}
\newcommand{\jwst}{\it JWST}
\newcommand{\kepler}{\it Kepler}
\newcommand{\kt}{\it K2}
\newcommand{\ktcn}{{\it K2}C9}
\newcommand{\spitzer}{\it Spitzer}
\newcommand{\swift}{\it Swift}
\newcommand{\wfirst}{\it WFIRST}

\begin{document}

\title{
On the Feasibility of Characterizing Free-floating Planets with Current and Future Space-based Microlensing Surveys
}

\author{
Calen B. Henderson\altaffilmark{1,A} and   
Yossi Shvartzvald\altaffilmark{1,A}   
}
\altaffiltext{1}{Jet Propulsion Laboratory, California Institute of Technology, 4800 Oak Grove Drive, Pasadena, CA 91109, USA}
\altaffiltext{A}{NASA Postdoctoral Program Fellow}
\email{calen.b.henderson@jpl.nasa.gov}

\begin{abstract}
Simultaneous space- and ground-based microlensing surveys, such as those planned with {\kt}'s Campaign 9 ({\ktcn}) and, potentially, {\wfirst}, facilitate measuring the masses and distances of free-floating planet (FFP) candidates.
FFPs are initially identified as events arising from a single lensing mass with a short timescale, ranging from one day for a Jupiter-mass planet to a few hours for an Earth-mass planet.
Measuring the mass of the lensing object requires measuring the angular Einstein radius {\thetaE}, typically by first determining the finite size of the source star $\rho$, as well as the microlens parallax {\piE}.
A planet that is gravitationally bound to, but widely separated from, a host star ($\gtrsim$20 AU) can produce a light curve that is similar to that of an FFP.
This tension can be resolved with high-resolution imaging of the microlensing target to search for the lens flux {\fl} from a possible host star.
Here we investigate the accessible parameter space for each of these components --- {\piE}, $\rho$, and {\fl} --- considering different satellite missions for a range of FFP masses, Galactic distances, and source star properties.
We find that at the beginning of {\ktcn}, when its projected separation from the Earth (as viewed from the center of its survey field) is $\lesssim$0.2 AU, it will be able to measure {\piE} for Jupiter-mass FFP candidates at distances larger than $\sim$2 kpc and to Earth-mass lenses at $\sim$8 kpc.
At the end of its campaign, when ${\dproj} = 0.81$ AU, it is sensitive to planetary-mass lenses for distances $\gtrsim$3.5 kpc, and even then only to those with mass $\gtrsim$M$_{\rm Jup}$.
From lens flux constraints we find that it will be possible to exclude \textit{all} stellar-mass host stars (down to the deuterium-burning limit) for events within $\sim$2 kpc, and for events at any distance it will be possible to exclude main sequence host stars more massive than $\sim$0.25 M$_{\odot}$.
Together these indicate that  the ability to characterize FFPs detected during {\ktcn} is optimized for events occurring toward the beginning of the campaign.
{\wfirst}, on the other hand, will be able to detect and characterize FFPs with masses at least as low as super-Earths throughout the Galaxy during its entire microlensing survey.

\end{abstract}


\section{{Introduction} \label{sec:intro}}

Understanding the frequency and mass function of free-floating planets (FFPs) is integral for a complete comprehension of the formation and evolution of planetary systems.
Using deep photometric imaging primarily in near-infrared bands and, in some cases, astrometric and/or spectroscopic follow-up data, Jupiter-mass FFPs and FFP candidates have been detected in the Trapezium cluster \citep{lucas2000}, the $\sigma$ Orionis open cluster \citep{bihain2009}, the AB Doradus moving group \citep{delorme2012,gagne2015}, the $\beta$ Pictoris moving group \citep{m_liu:2013,allers2016}, the Upper Scorpius Association \citep{pena_ramirez2015}, and in the field \citep{dupuy2013}.
Yet detections from photometric surveys require and depend sensitively on an independent constraint for the object's age while the dynamical association of objects in moving groups is intrinsically more uncertain.

In a broader statistical sense, \citet{sumi2011} examined two years of microlensing survey data from the Microlensing Observations in Astrophysics (MOA) collaboration.
They found an excess of short-timescale events ($<$2 days) above expectations based on an extrapolation of the stellar mass function down to low-mass brown dwarfs.
From this they inferred a population of FFP candidates that outnumber main sequence stars by a ratio of $1.8^{+1.7}_{-0.8}$.
However, mass measurements do not exist for any of the candidate objects upon which the inference rests.

The formation mechanisms for FFPs remain an open theoretical question.
One possible avenue is that these objects were originally formed in protoplanetary disks and were subsequently ejected.
Simulations by \citet{pfyffer2015} of the formation and evolution of planetary systems without eccentricity or inclination damping eject planets at a rate that is $\gtrsim$50 times lower than is needed to explain the MOA result.
Indeed, \citet{veras2012} assert that planet-planet scattering itself is insufficient to reproduce the abundance of FFP candidates seen by MOA.
In the context of open clusters, N-body dynamical simulations find that roughly half of star-planet and planet-planet interactions occur within the first 30 Myr \citep{wang2015} and that 80$\%$ of the resulting FFPs are ejected from the cluster while the remaining $\sim$20$\%$ become concentrated in the central $\sim$2 pc \citep{h_liu:2013}.

A second option is that FFPs form via direct collapse of molecular clouds.
\citet{silk1977} found that opacity-limited fragmentation of collapsing clouds could, in principle, produce a fragment with a minimum mass as low as $\sim$0.01${\rm M}_{\odot}$.
Additionally, turbulent shocks could cause protostellar cores with masses in the brown dwarf regime to become gravitationally unstable and hypothesized that turbulent density fluctuations smaller than a critical mass could induce collapse into objects with the mass of giant planets \citep{padoan2004}.

Gravitational microlensing does not rely on the flux output from the lensing object and so is well-positioned to explore FFP demographics across a wide range of planet masses.
The mass of an FFP lens, {\mffp}, is given by:
\begin{equation} \label{eq:mffp}
   {\mffp} = {\thetaE}/(\kappa{\piE}),
\end{equation}
where {\thetaE} is the angular Einstein radius, $\kappa \equiv 4G/(c^{2}{\rm AU}) = 8.144~{\rm mas}/M_{\odot}$, and {\piE} is the microlens parallax:
\begin{equation} \label{eq:piE}
   {\piE} = {\rm AU}/{\rEproj} = {\pirel}/{\thetaE}.
\end{equation}
Here {\pirel} is the lens-source relative parallax, defined as ${\rm AU}({\dl}^{-1} - {\ds}^{-1})$, where {\dl} and {\ds} are the distances from the observer to the lens and source, respectively, and {\rEproj} is the physical size of {\thetaE} projected onto the plane of the observer.

The most salient observable for a light curve arising from a single lensing mass is the Einstein timescale {\tE}, which measures the time required for the lens-source angular separation to change by one angular Einstein ring radius and is defined as:
\begin{align} \label{eq:tE}
   {\tE} & \equiv \frac{\thetaE}{\murel} \\
   & \simeq 1.4~{\rm hr} \left(\frac{\mffp}{M_{\oplus}}\right)^{1/2} \left(\frac{\pirel}{125~\mu{\rm as}}\right)^{1/2} \left(\frac{\murel}{10~{\rm mas~yr^{-1}}}\right)^{-1}, \nonumber
\end{align}
where {\murel} is the lens-source relative proper motion.
The timescale is routinely measured for single-lens events.
However, several physical properties are encoded within {\tE}, including {\dl}, {\mffp}, and {\murel}.
Thus, in order to determine that the lensing object giving rise to the short timescale is indeed planetary-mass, it is crucial to measure both {\thetaE} and {\piE}.
The angular Einstein radius is typically determined by combining a measurement of $\rho$, the angular radius of the source star normalized to {\thetaE}, with multiband photometry to determine the color and, ultimately, the angular size of the source \citep{yoo2004}.
Lastly, a short-timescale event definitively due to a planetary-mass lens must be proven to be free-floating, which requires high-resolution photometry to search for flux from a possible lens host star.
\citet{han2005b} studied bound planets that are widely separated from their host star.
They found that if the projected separation between a planet and its parent star is $\gtrsim$20 AU, the resulting light curve can mimic that of an FFP, as the alignment between the source trajectory and the binary lens axis can cause the primary microlensing event due to the star \textit{not} to be observed, allowing a bound planet to masquerade as an FFP.
They also investigated the ability to identify the bound nature of a planet in the case of isolated planetary-mass lensing events via the caustic induced by the shear of the planet's host star, which would introduce magnification structure to the light curve of the event.

Fully characterizing an FFP requires that all of these criteria are met.
However, the intersection of all aforementioned constraints is far from guaranteed for a given space mission.
The goal of this paper is to map the regions of parameter space for which a given satellite will be sensitive to FFPs, particularly {\kt}'s Campaign 9 ({\ktcn}; see \S \ref{sec:kepler}) and {\wfirst} (see \S \ref{sec:wfirst}), and to identify the dominant limiting factors in each regime.
Throughout the paper we will assume that the source is located in the bulge at a distance of ${\ds}=8.2$ kpc \citep{nataf2013} and focus on {\dl} as an independent variable.
It is possible to explore the demographics of FFPs statistically without requiring that the parameters {\piE}, $\rho$, and {\fl} each be measured for all systems.
That said, the primary focus of this paper is on the subset of FFPs that can be fully characterized via measurements of all three.

The expected yield of FFPs for a given microlensing survey depends, intrinsically and sensitively, on their underlying event rate.
Understanding the relative rates for different FFP populations can influence the observational strategy.
The microlensing rate as a function of the FFP physical parameters, derived from the generic rate formula $\Gamma=n\sigma v$, is given by \citep{batista2011}:
\begin{equation} \label{eq:rate}
   \frac{{\partial}^{4}\Gamma}{{\partial}{\dl} {\partial}{\mffp} {\partial}^{2} \mu} = n(x,y,z) (2{\rE}) v_{\rm rel} f(\mu)  g({\mffp}).
\end{equation}
Here $n(x,y,z)$ is the local space density of FFPs, $f(\mu)$ is the lens-source relative proper motion probability distribution, and $g({\mffp})$ is the FFP mass function.
While {\rE} is proportion to ${\mffp}^{1/2}$ and $f({\mu})$ can be assumed to be similar to that for the general Galactic stellar population, $n(x,y,z)$ and $g({\mffp})$ are precisely what the experiment is trying to measure and are hitherto unconstrained.
Furthermore, the kinematics for FFPs will depend on the fraction that are the result of dynamical ejections versus those that form in situ.
Thus, in this paper we cannot and do not predict the relative rates for different FFP populations but instead study the feasibility of characterizing FFPs with various physical parameters, assuming that an event has occurred.

Our methodology builds upon and extends previous studies of lens characterization for FFPs and also bound planets.
\citet{han2004} examined the ability of a joint ground- and space-based microlensing survey to constrain the mass of FFPs.
Considering a generic satellite at the Earth-Sun L2 point paired with a ground-based survey, they found that microlensing events arising from planetary-mass lenses can be detected for planet masses down to that of the Earth.
\citet{han2006b} subsequently explored how to distinguish between isolated lensing events due to widely separated, bound planets from FFP events using interferometry to search for the astrometric signature of a lens host star.
He found that for bright sources, with $V \lesssim 19$, a centroid shift can be measured given an astrometric precision that is of-order one microarcsecond.
More generally, \citet{jyee:2015} investigated how to measure lens masses and distances solely from measurements of the microlens parallax {\piE} and the lens flux {\fl}, circumventing the need to measure {\thetaE}.
She considered spacecraft in Earth-trailing Solar orbits and at Earth-Sun L2, similar to those of {\kepler}, {\spitzer}, and {\wfirst}.
Naturally, lens flux cannot be used as a tool for characterizing FFPs, necessitating an additional constraint.
\citet{zhu2016} advocate for a simultaneous ground-based survey to accompany the {\wfirst} microlensing observations in order to obtain two-dimensional vector microlens parallax measurements.
These would provide complete solutions, and thus masses and distances, for a substantial fraction of FFP events.
We present the first combined treatment of constraints from {\piE}, $\rho$, and {\fl} to explore regimes of detectability and characterization for FFPs across a range of lens distances, source star properties, and satellite characteristics.

In \S \ref{sec:satellites} we briefly describe the parameters and goals of six relevant space missions.
Measuring {\piE} with a space telescope requires a delicate balance between {\rEproj} and the projected separation {\dproj} of the Earth and the satellite as seen from the lens-source line-of-sight.
We investigate the interplay between {\dproj} and {\rEproj} for these satellite missions in \S \ref{sec:dproj_rEproj}.
In \S \ref{sec:rho} we discuss the probability of measuring $\rho$ through the detection of finite-source effects.
Finally, in \S \ref{sec:fl} we detail the constraints from measuring the lens flux.
We summarize our findings and discuss their applications in the context of {\ktcn} and {\wfirst} in \S \ref{sec:discussion}, wherein we identify the regimes of parameter space to which each will ultimately be able to robustly characterize FFP events.

\section{Overview of Satellite Missions} \label{sec:satellites}

In this section we provide an overview of all space telescopes that are potentially relevant for monitoring microlensing events simultaneously with ground-based resources.

\subsection{{\spitzer}} \label{sec:spitzer}

{\spitzer}, a 0.85m infrared telescope, was the first satellite used to conduct real-time monitoring of a microlensing event simultaneous with ground-based facilities \citep{dong2007}.
A 100-hour campaign in 2014 expanded upon this in an effort to make such measurements systematically for multiple events.
The 2014 program led to the first satellite parallax measurement for an isolated star \citep{yee2015b} and a microlensing-discovered exoplanet \citep{udalski2015b}.
A larger 832-hour program followed in 2015, for which target events were selected using objective criteria in order to maximize planet sensitivity and number of detections \citep{yee2015a}.
The resulting discoveries include mass and distance measurements for a cold Neptune in the Galactic disk \citep{street2015}, a massive remnant in a stellar binary \citep{shvartzvald2015}, and several isolated objects, including a brown dwarf \citep{zhu2015a}, demonstrating the wide range of astrophysical populations that can be probed with such a technique. These campaigns are the first steps for measuring the Galactic distribution of bound planets \citep{calchinovati2015a}.

Two additional {\spitzer} programs will take place in 2016, one to explore the Galactic distribution of exoplanets using high-magnification microlensing events \citep{gould2015b} and the other to conduct a two-satellite microlensing experiment \citep{gould2015a} by observing in conjunction with {\ktcn} (see \S \ref{sec:kepler}).
However, the short-timescale events that are indicative of FFPs are generally inaccessible with {\spitzer} due to the several-day lag between target selection and upload and the first possible observations (see Figure 1 of \citealt{udalski2015b}).
Nevertheless, we include it here because it is possible to have a combination of physical parameters {\dl}, {\mffp}, and {\murel} that produces a longer timescale for an FFP, and also for the sake of completeness.

\subsection{{\kepler}} \label{sec:kepler}

{\kepler} is 0.95m telescope in an Earth-trailing Solar orbit whose primary mission was to explore exoplanet demographics using the transit method.
The mechanical failure of the second of its four reaction wheels in 2013 signaled an end to the primary mission but heralded the genesis of its extended {\kt} Mission, which is in the midst of a series of $\sim$80-day campaigns performing high-precision photometry for targets along the Ecliptic \citep{howell2014}.
{\ktcn} will conduct the first microlensing survey from the ground and from space, covering 3.74 deg$^{2}$ from 7/April through 1/July of 2016 in concert with a vast array of ground-based resources \citep{sexypants2015b}.
Approximately $\gtrsim$120 events will occur in the {\ktcn} microlensing survey field (termed ``superstamp") during the campaign dates, of-order 10 of which are estimated to be short-timescale FFP candidates.

{\kt}'s Campaign 11 will also point toward the Galactic bulge from 24/September through 8/December of 2016.
While it will not perform an automated microlensing survey, it will yet be possible to measure {\piE} for events detected by ground-based telescopes that will have timescales long enough that they will be ongoing when {\kt} observations take place, though this excludes short-timescale FFP candidates.

\subsection{{\wfirst}} \label{sec:wfirst}

{\wfirst}, which was recently approved for Phase A development, will be a 2.4m telescope equipped with a wide-field imager with a field of view of 0.28 deg$^{2}$ and six imaging filters that span 0.76--2.0 microns \citep{spergel2015}.
Set to launch in the mid-2020s, it will be inserted into a Solar orbit at Earth-Sun L2.
{\wfirst} will conduct a $\sim$432-day microlensing survey toward the Galactic bulge divided equally between six 72-day seasons.
These seasons will be split between the beginning and end of the mission to maximize the ability of {\wfirst} to measure {\murel} for the detected events.
Furthermore, its orbital placement and observational parameters will make it the most advanced experiment in terms of FFP detection and characterization.

\subsection{{\euclid}} \label{sec:euclid}

The scientific focus of {\euclid} will be to explore the nature of dark energy.
However, an exoplanetary microlensing survey could be conducted as part of the mission's legacy program and would, in principle, be sensitive to FFPs \citep{penny2013}.
{\euclid}, which is scheduled to launch in 2020, will be placed in a halo orbit at Earth-Sun L2.
So, while its 1.2m aperture is smaller than that of {\wfirst}'s, the orbital configuration and resulting geometric sensitivity will be quite similar to that of {\wfirst}, which we derive here.
As such, we do not explicitly consider {\euclid} as a separate mission throughout the remainder of the text and instead note that {\wfirst} is a good proxy for its capabilities.

\subsection{{\swift}} \label{sec:swift}

During the 2015 {\spitzer} campaign, the microlensing event OGLE-2015-BLG-1395 (OB151395) was observed with {\swift}, a 0.3m telescope that can observe from optical to $\gamma$-ray, while it was highly magnified, with $V \lesssim 16.5$.
OB151395 is thus the first event with observations from two satellites while the event was ongoing (Shvartzvald et al., in prep.).
While the rapid response of {\swift} is ideal for short-timescale FFPs, it is uncommon to have events reach a sufficiently bright magnitude to be detectable for the Ultra Violet/Optical Telescope.
Furthermore, {\swift} is on a low-Earth orbit, so its distance from the surface of the Earth is within the range $\sim$560--576 km, leading to an extremely small projected separation (see \S \ref{sec:dproj}) that moreover varies significantly during its $\sim$96-minute orbital period.
That said, {\swift}'s orbit can be advantageous for events sensitive to terrestrial parallax.

\subsection{{\newhorizons}} \label{sec:newhorizons}

{\newhorizons} completed a six-month flyby of Pluto in the summer of 2015 and is continuing into the outer reaches of the Solar System to study the Kuiper belt.
Its projected separation is, by a factor of several, the largest of the satellites we will discuss here.
While this indicates that it will be all but impossible for {\newhorizons} to detect FFPs, we include it as an upper limit and note that it is potentially well-suited to measure {\piE} for lens systems that are much more massive than FFPs and generally have longer timescales and larger Einstein radii, such as stellar remnants.

\section{Measuring the Microlens Parallax {\piE}} \label{sec:dproj_rEproj}

A microlensing event due to a single lensing mass leads to a light curve defined by three microlensing observables \citep{paczynski1986}.
The first is {\tzero}, the time of closest approach of the source to the lens.
Second is the impact parameter {\uzero}, which measures the angular distance of the closest approach of the source to the lens and is normalized to {\thetaE}.
Third is the timescale {\tE}, defined in Equation (\ref{eq:tE}).
Measuring the microlensing parallax {\piE} for a short-timescale microlensing event requires measuring the shift in {\tzero}, {\uzero}, or both, in the light curve as it is seen from two (or more) locations (for a recent review on this method see \citealt{calchinovati2015b}).
Together these observables define the ratio of {\dproj}, the projected separation between two observers, to {\rEproj} via:
\begin{equation} \label{eq:dproj_rEproj}
   \frac{\dprojvec}{\rEproj} = (\frac{\Delta{\tzero}}{\tE},\Delta{\uzero}).
\end{equation}
In principle, {\piE} can also be measured by a single, accelerated observer for long-timescale events, for which {\tE} is of-order the orbital period (i.e., ${\tE} \gtrsim {100}$ d), though this is not relevant for FFPs.

\subsection{Satellite-Earth Projected Separation {\dproj}} \label{sec:dproj}

\begin{figure}
   \centerline{
      \includegraphics[width=9cm]{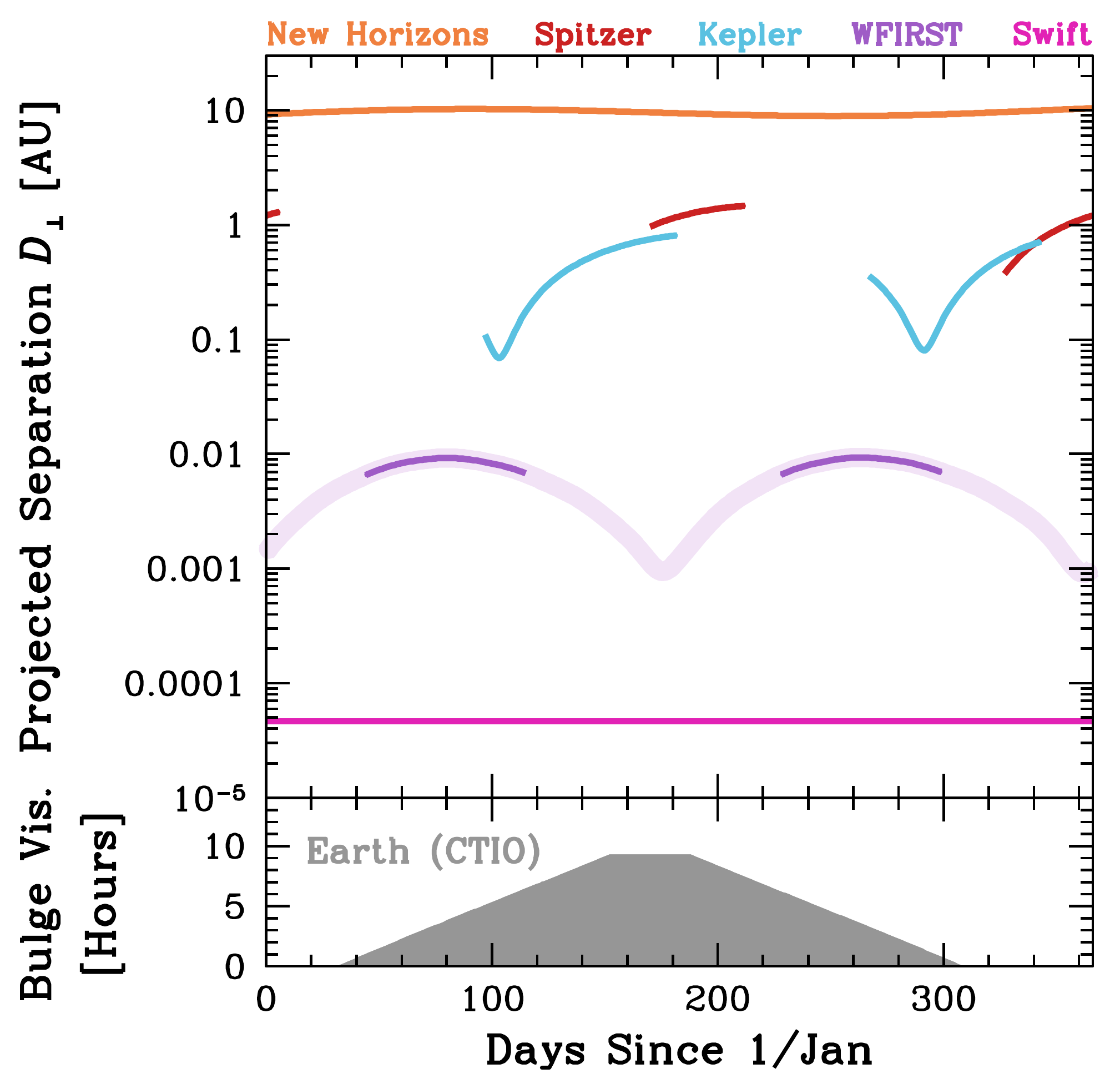}
   }
   \caption{
      \footnotesize{
         The projected separation {\dproj} between the Earth and several satellite missions (see \S \ref{sec:satellites}) throughout a generic year (top).
         In the bottom panel we show the number of hours that the Galactic bulge is visible from CTIO during the ground-based bulge observing season.
         Regarding {\spitzer}, the red curves identify when the viewing angle $\psi$ between the Sun and the bulge, for which we use the approximate center of the {\ktcn} superstamp: (RA, Dec) = $(17^{\rm h}56^{\rm m}54^{\rm s}, -28^{\rm d}22^{\rm m}5^{\rm s})$, satisfies: $82.5 \le \psi/{\rm deg} \le 120$.
         The two blue lines denote when {\kepler} will be oriented toward the bulge: during its Campaign 9 (left) and Campaign 11 (right).
         {\wfirst}, for which we use the orbit of {\gaia} as a proxy (see \S \ref{sec:dproj} for discussion), will not observe the bulge continuously but instead during six 72-day microlensing seasons \citep{spergel2015}.
         The exact dates of these seasons have yet to be set but will likely be centered on the equinoxes due to Sun angle pointing restrictions.
         Thus, the broad light purple band shows the full orbit while the thinner dark purple curves show a 72-day window centered on each equinox.
         For {\swift}, in pink, we show the maximum {\dproj} it attains over the course of its $\sim$96-minute low-Earth orbit and note that this value is computed from the Earth's geocenter, indicating that its projected separation from an observatory on Earth can differ, potentially significantly.
         For {\newhorizons}, the annual parallactic oscillation in its projected separation is primarily due to the orbital motion of the Earth.
         All orbits were computed using the JPL \texttt{HORIZONS} web interface (see \S \ref{sec:dproj}).
         \vspace{1.0mm}
      }
   }
   \label{fig:dproj}
\end{figure}

Figure \ref{fig:dproj} shows the projected separation {\dproj} between Earth and each of the six satellite missions discussed in \S \ref{sec:satellites} as seen from the center of the {\ktcn} superstamp throughout a generic year.
In all cases we compute orbits using the JPL \texttt{HORIZONS} web interface\footnote{\url{http://ssd.jpl.nasa.gov/horizons.cgi}}.
We use the center of the {\ktcn} superstamp, (RA, Dec) = $(17^{\rm h}56^{\rm m}54^{\rm s}, -28^{\rm d}22^{\rm m}5^{\rm s})$, as the approximate center of bulge observations, coupled with the Earth's geocenter when computing {\dproj} for all satellites.

Over the course of {\ktcn} its projected separation will span $0.07 \lesssim {\dproj}/{\rm AU} \lesssim 0.81$.
The visibility windows for {\spitzer} are set by the requirement that the viewing angle $\psi$ between the Sun and the target falls within the range: $82.5 \le \psi/{\rm deg} \le 120$.
This highlights the synergy between {\kepler} and {\spitzer} that will occur during the last two weeks of {\ktcn}, though it is unlikely that {\spitzer} will obtain data for any FFP candidates.
For {\swift}, we show the maximum {\dproj} it reaches over the course of its $\sim$96-minute low-Earth orbit.
We note that its projected separation varies significantly over the course of its orbit and also as a displacement measured from any observatory on the surface of the Earth.
For {\newhorizons}, the annual parallactic oscillation in its projected separation is primarily due to the orbital motion of the Earth.

Lastly, {\wfirst} will not observe the bulge continuously but will feature six 72-day observing seasons dedicated to microlensing.
The most recent Sun angle pointing requirements for the planned spacecraft indicate that the microlensing seasons will be centered on the equinoxes.
This indicates that {\dproj} will be near its maximum for much of these seasons and also that simultaneous ground-based observations will be restricted due to the visibility of the Galactic bulge.
The {\wfirst} spacecraft will be inserted into the Earth-Sun L2 Lagrange point.
Due to the dynamically unstable nature of L2, {\wfirst} will trace out a halo orbit around the L2 point.
Currently no orbital data exist on JPL's \texttt{HORIZONS} for {\wfirst}, so we investigate the orbits of two other L2 spacecraft as proxies.
{\jwst} will also be inserted into a halo orbit at Earth-Sun L2, but its periapsis and apoapsis will each be larger than that of {\wfirst} by a factor of $\sim$2.
{\gaia} is on a Lissajous orbit around L2 with a periapsis and apoapsis that are each within a few tens of percent of that of {\wfirst}.
While halo orbits are periodic and Lissajous orbits are quasi-periodic, the L2 orbits of {\jwst}, {\wfirst}, and {\gaia} all feature a periodicity of $\sim$180 days.
We thus use the orbit of {\gaia} as a proxy for that of {\wfirst} and note that the morphology and extrema for the final true orbit of {\wfirst} will be slightly altered from what is presented here.

\subsection{Projected Physical Einstein Radius {\rEproj}} \label{sec:rEproj}

\begin{figure}
   \centerline{
      \includegraphics[width=9cm]{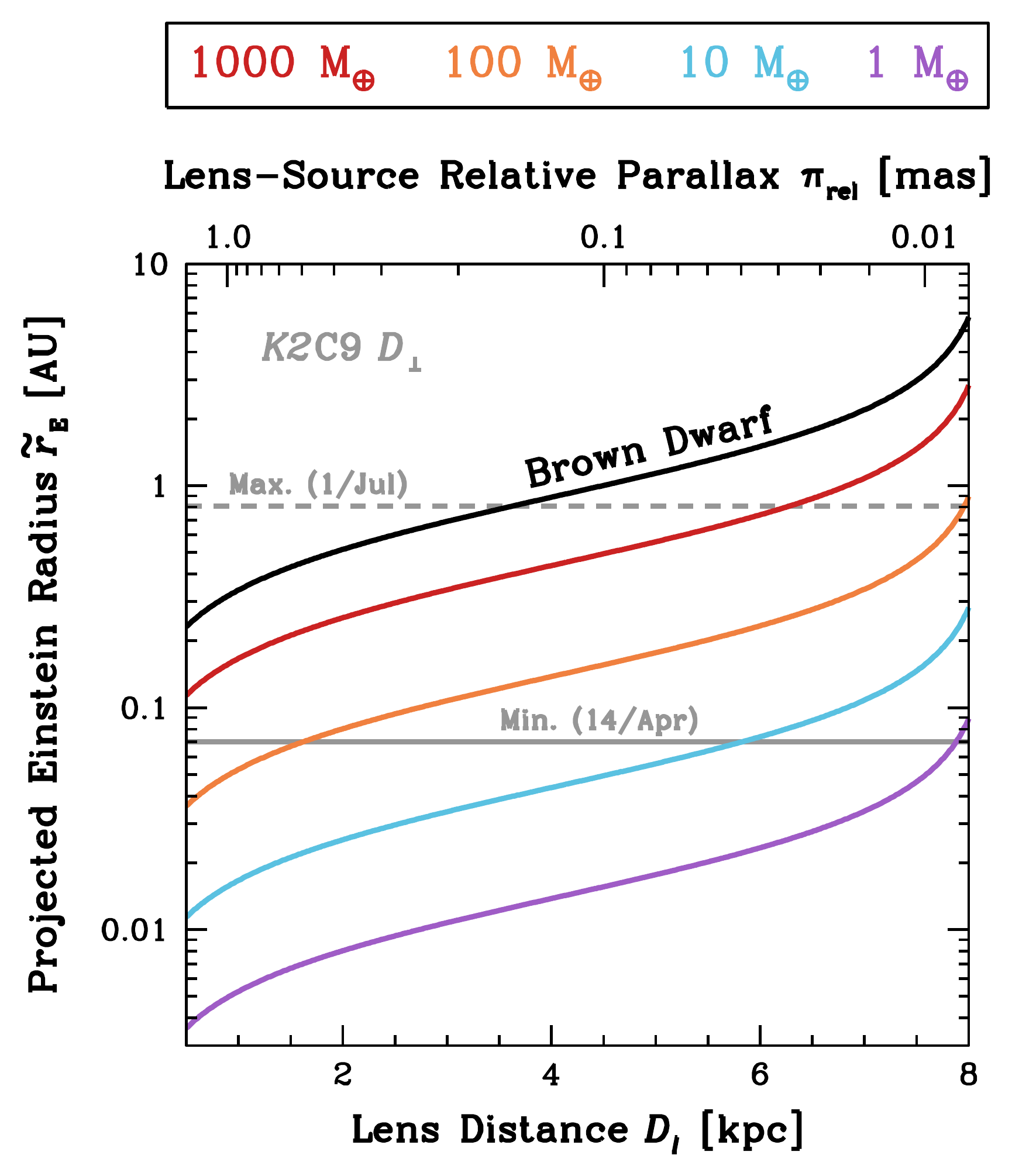}
   }
   \caption{
      \footnotesize{
         The physical size of the Einstein radius projected onto the observer plane, {\rEproj}, as a function of lens distance {\dl} (bottom axis) and lens-source relative parallax {\pirel} (top axis), for several planet masses {\mffp} and for the brown dwarf mass limit ($\sim$4100 {\mearth}).
         For comparison, the horizontal grey lines indicate the time of the minimum (0.07 AU) and maximum (0.81 AU) values of {\dproj} for {\ktcn}.
         This is not a strict cutoff for detection of FFP candidate events, as discussed in \S \ref{sec:rEproj}, but provides a reasonable estimate of the combinations of {\mffp} and {\dl} that will be detectable with {\ktcn} throughout its campaign.
         \vspace{1.0mm}
      }
   }
   \label{fig:rEproj}
\end{figure}

The projected physical size of the Einstein ring is related to the physical parameters {\mffp} and {\pirel} by:
\begin{equation} \label{eq:rEproj}
   \rEproj=\sqrt{\frac{\kappa{\mffp}}{\pirel}}{\rm AU}.
\end{equation}
Figure \ref{fig:rEproj} shows {\rEproj} for several planet masses as well as the brown dwarf mass limit (13 {\mjup}, $\sim$4100 ${\rm M}_{\oplus}$), as a function of {\dl}.

We consider a microlensing event to be detectable if $|{\uzero}| \le 1$ from both the ground and space.
In principle, though, it is possible to detect an event with $|{\uzero}| > 1$ if the source is very bright.
Defining $\phi$ as the angle between the lens-source relative proper motion {\murelvec} and the Earth-satellite projected separation {\dproj}, we establish the following relation:
\begin{equation} \label{eq:delta_uzero}
   \frac{\dproj}{\rEproj}{\rm sin}\phi = \Delta {\uzero}.
\end{equation}
The criterion for detectability, when coupled with Equation (\ref{eq:delta_uzero}), means that even if $|{\uzero}| \le 1$ from one observatory, there will exist values of $\Delta {\uzero}$ for which $|{\uzero}| > 1$ for the second observatory, leading to an undetectable event.
Furthermore, the existence of a preferred direction for {\murelvec} will necessarily select a locus of $\phi$ and consequentially, via Equation (\ref{eq:delta_uzero}), a range of $\Delta {\uzero}$.
As a result, even for an assumed distribution of {\uzero} for the first observatory that is uniform, the distribution of {\uzero} for the second observatory would \textit{not} be uniform and, as mentioned above, would include a population of undetectable events.
A lower limit of $\Delta {\uzero}$, which is implied in the case of a non-detection of the event by one observer, places a lower limit on {\piE} and hence an upper limit on the lens mass, if {\thetaE} is measured.
Given this, we note that even with a sample of events that are detected only from the ground \textit{or} from space, and for which there exists only a lower limit on {\piE}, it is still possible to statistically explore the FFP mass function.

In the Milky Way there is a preferred direction of {\murelvec} defined by the direction of Galactic rotation, so here we determine the probability that an event will be detectable as a function of ${\dproj}/{\rEproj}$.
The probability distribution of the proper motion {\murelvec} is a two-dimensional (Galactic coordinates: $l,b$) Gaussian with expectation value:
\begin{equation} \label{eq:murel_expect}
   <{\murelvec}>~=~\frac{{\vl}-{\vsun}}{\dl} - \frac{{\vs}-{\vsun}}{\ds},
\end{equation}
where {\vl}, {\vs}, and {\vsun} are the velocity of the lens, source, and Sun, in Galactic coordinates.
We model each as a Gaussian and list the mean and standard deviation for each object in Table \ref{tab:v_dist}.
Since all satellites considered here are on the Ecliptic and the bulge microlensing fields are generally close to the Ecliptic ($\lesssim$5$^\circ$), we rotate the coordinates by 60$^\circ$ (the angle between the Ecliptic and the Galactic plane) and derive $\phi$.
The correction for fields off of, but close to, the Ecliptic under this approximation should thus be $\lesssim$5$^\circ$.
Moreover, events due to lenses in the bulge have a negligible preferred direction (see Table \ref{tab:v_dist}), so this sets an upper limit for the error on our assumed rotation.

We note that the preferred direction of {\murelvec} also suggests a preferred sign of $\Delta{\tzero}$, primarily for disk lenses.
On average, events will be detected first by a western observer, e.g., an Earth-trailing satellite.
This can have implications when considering events alerted from one observer location and followed up by a second observer.
For {\wfirst} this would indicate a shift in the origin of first-detection alerts for a season centered on the vernal equinox compared to a season centered on the autumnal equinox.

\begin{deluxetable}{ccccc}
\tablecaption{Velocity Distributions$^{a}$}
\tablewidth{0pt}
\tablehead{
\colhead{Object}             &
\colhead{$v_{b}$}            &
\colhead{$v_{l}$}            &
\colhead{$\sigma_{v_{b}}$}   &
\colhead{$\sigma_{v_{l}}$}
}
\startdata
Disk lens    &   0   &   220   &   20    &   30    \\
Bulge lens   &   0   &   0     &   100   &   100   \\
Source               &   0   &   0     &   100   &   100   \\
Sun                  &   7   &   242   &   0     &   0
\enddata
\tablenotetext{a}{All values are in km~s$^{-1}$.}
\label{tab:v_dist}
\end{deluxetable}

With the velocity distributions in hand, we compute the probability of a detectable event, i.e., an event that is seen from both the Earth and a satellite.
We first draw {\uzero} for one observer (the order of observatories is immaterial in this context) from a probability distribution of $P(|{\uzero}| \le 1)$ (i.e., detected by the first observer).
Next, we draw a direction from the proper motion probability distribution for lenses in the disk with {\dl}  = 1 and 4 kpc and bulge lenses with {\dl} = 7 kpc.
Using Equation (\ref{eq:delta_uzero}) we then derive $\Delta {\uzero}$ for a range of {\dproj}/{\rEproj}, specifically, $0.1 \le {\dproj}/{\rEproj} \le 10$.
Finally, after adding $\Delta {\uzero}$ to the {\uzero} of the first observer, we check if $|{\uzero}| \le 1$ also for the second observer.

In principle, $P(|{\uzero}| \le 1)$ should be uniform.
However, as \cite{shvartzvald2012} and \cite{shvartzvald2016} showed, there is a selection effect that, for intrinsically faint sources, favors the detection of high-magnification events (i.e., small {\uzero}) due to the limiting magnitudes of the surveys.
This bias will persist for current ground and space surveys.
In the case of {\wfirst}, most of the sources will be bright enough to be detected at baseline from space and, potentially, from the ground as well, though the resources contributing to a simultaneous ground-based survey are currently unknown.
Additionally, since, for a significant fraction of the FFP parameter space, the maximum possible magnification is relatively low (see \S \ref{sec:rho}), this bias will be less prevalent.
Therefore, we calculate the results both for a uniform distribution of {\uzero}
and for the observed distribution found by \cite{shvartzvald2012} and \cite{shvartzvald2016}, which favors small {\uzero}.

\begin{figure}
   \centerline{
      \includegraphics[width=9cm]{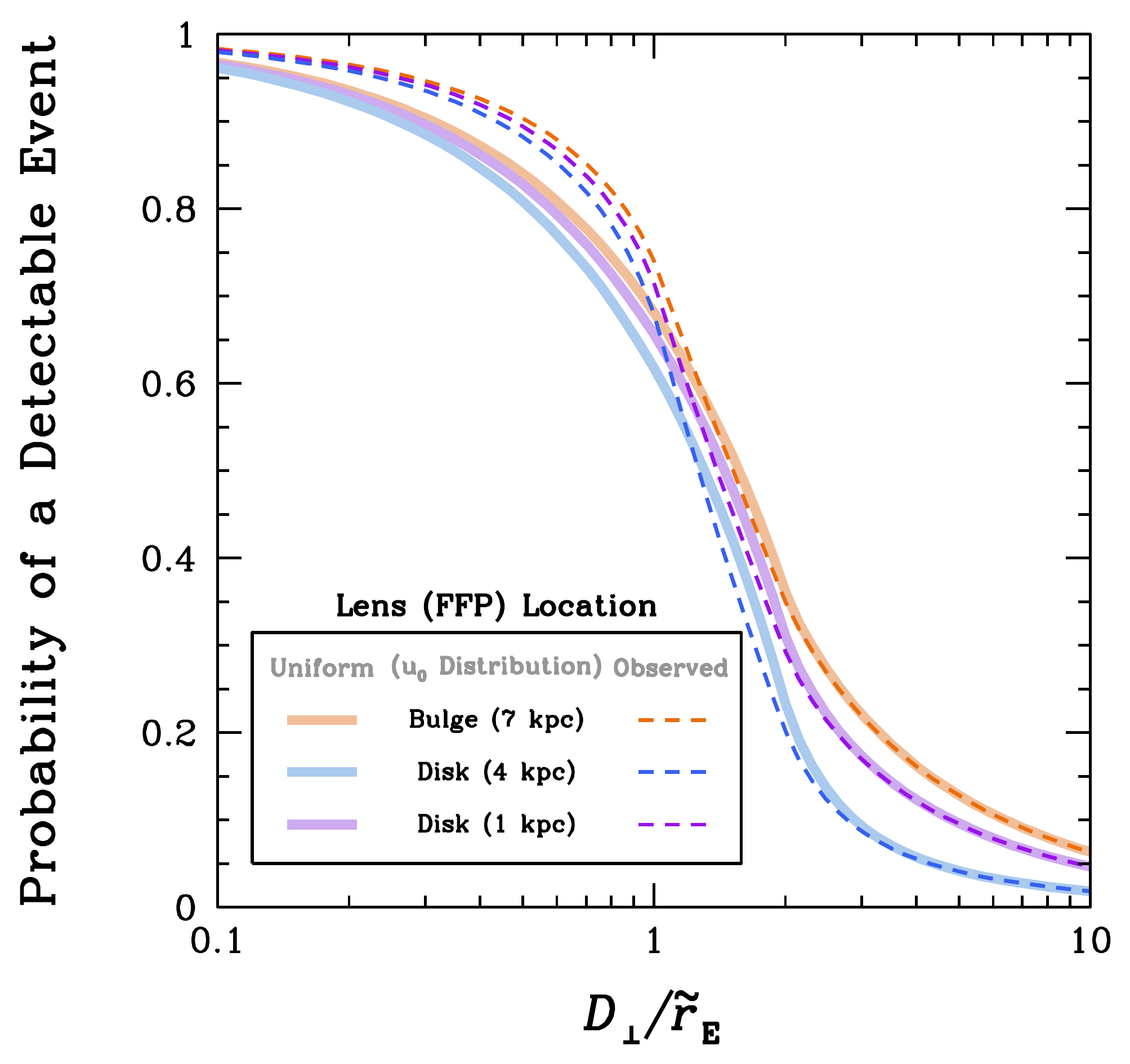}
   }
   \caption{
      \footnotesize{
        The probability for an event to be detected from both the ground and a satellite (i.e., $|{\uzero}| \le 1$ for both) as a function of {\dproj}/{\rEproj} for a lens (e.g., FFP) in the bulge with ${\dl} = 7$ kpc (orange), in the disk with ${\dl} = 4$ kpc (blue), and in the disk with ${\dl} = 1$ kpc (purple).
        For {\uzero} we use both a uniform distribution (solid lines) and the observed distribution (dashed lines; \citealt{shvartzvald2012,shvartzvald2016}), which favors smaller {\uzero}, and compute the probability as described in \S \ref{sec:rEproj}.
        The probability is $>$60$\%$ for ${\dproj} < {\rEproj}$, 20--36$\%$ for ${\dproj} = 2{\rEproj}$, and it decreases monotonically for ${\dproj} > 2{\rEproj}$.
        \vspace{1.0mm}
      }
   }
   \label{fig:prob_dproj_rEproj}
\end{figure}

Figure \ref{fig:prob_dproj_rEproj} shows the probability of a detectable event (i.e., $|{\uzero}| \le 1$ from both observer locations) as a function of {\dproj}/{\rEproj} for the three lens distances and two {\uzero} distributions we explore.
For ${\dproj} < {\rEproj}$, the probability is $>$60$\%$ for all possibilities, with the observed {\uzero} distributions yielding a higher probability by up to 8$\%$ as expected.
For ${\dproj} = 2{\rEproj}$ the probability falls to 20--36$\%$, with a higher probability for the uniform distribution by $\sim$2$\%$.
This arises from the fact that on average $\Delta{\uzero} > 1$, so a higher probability of having a smaller {\uzero} for one observer suggests that for the other $|{\uzero}|>1$.
Beyond ${\dproj} > 2{\rEproj}$ the probability declines gradually.

\section{Normalized Angular Source Star Radius $\rho$} \label{sec:rho}

An additional requirement for a secure FFP characterization is the determination of {\thetaE}, which must be done via the measurement of finite-source effects, given the absence of detectable lens flux and high-precision astrometric data.
The geometric probability $P$ that the projected separation between the source and the FFP lens  will be sufficiently small that the angular size of the source is measurable can be cast in terms of the physical parameters as:
\begin{equation} \label{eq:rho_prob}
   P\simeq\rho=\frac{\thetastar}{\sqrt{\kappa{\mffp}{\pirel}}},
\end{equation}
where {\thetastar} is the angular radius of the source star.
In general, it is necessary that ${\uzero} \lesssim \rho$, for at least one of the observers, in order for the finite angular size of the source to be measured.
We note that for $\rho \lesssim 0.01$, the probability can be boosted by a factor of $\sim$2 for two observers with ${\dproj} < {\rEproj}$, assuming a uniform distribution for {\uzero}, and up to a factor of $\sim$4, given the observed {\uzero} distribution.
However, as $\rho$ increases, the probability converges to Equation (\ref{eq:rho_prob}).

\begin{figure*}
   \centerline{
      \includegraphics[width=18cm]{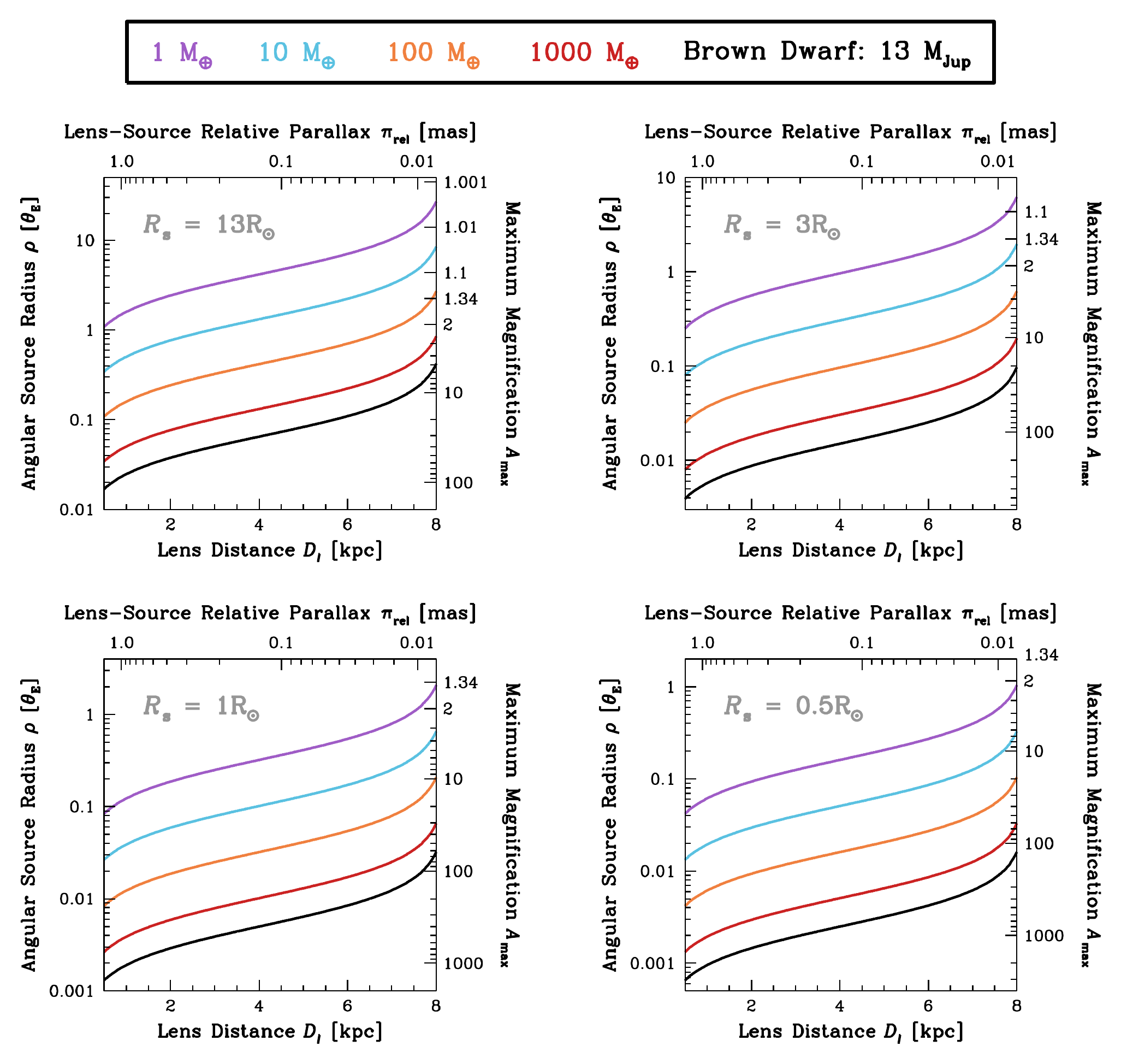}
   }
   \caption{
      \footnotesize{
         The angular size of the source star normalized to the Einstein radius, $\rho$, as a function of lens distance {\dl} (lower x-axis) and lens-source relative parallax {\pirel} (upper x-axis) for several planet masses {\mffp}.
         Each panel represents a source star with a different physical radius for four typical spectral types: a giant ({\rs} = 13{\rsun}; top left), a subgiant (3{\rsun}; top right), a Solar-type main sequence star (1{\rsun}; bottom left), and a late K/early M dwarf (0.5{\rsun}; bottom right).
         We assume {\ds} = 8.2 kpc \citep{nataf2013}, with which we compute the angular radius of the source star {\thetastar} for all cases.
         The right-hand y-axis gives  the maximum possible magnification corresponding to each $\rho$, which is significantly suppressed for a giant source, while the probability of measuring finite-source effects, and thus $\rho$, for individual events is much lower for main sequence sources.
         \vspace{1.0mm}
      }
   }
   \label{fig:rho}
\end{figure*}

Figure \ref{fig:rho} shows $\rho$ as a function of {\dl} for several FFP masses.
Each panel represents a typical physical source radius {\rs} for different source spectral types: {\rs} = 13{\rsun} for a clump giant, 3{\rsun} for a subgiant, 1{\rsun} for a main sequence Solar-type source, and 0.5{\rsun} for a late K/early M dwarf.
It is crucial to note that the probability of detecting finite-source effects computed in Equation (\ref{eq:rho_prob}) is counterbalanced by the fact that the maximum possible magnification decreases for larger $\rho$.
\cite{witt1994} found that the maximum magnification {\amax} as a function of $\rho$ is:
\begin{equation} \label{eq:rho_amax}
   {\amax} = \sqrt{\rho^2 + 4} / \rho.
\end{equation}
The right-hand y-axis of Figure \ref{fig:rho} provides the values of {\amax}.
As $\rho$ increases, not only is the maximum possible magnification suppressed, but the light curve broadens.
Consequentially, the fractional deviation of the wings of the light curve from that expected for a point-like source can be substantial.
However, it becomes increasingly possible to reproduce the light curve with a range of impact parameters, a degeneracy that ultimately impacts the ability to robustly measure {\piE}.

For giant sources, the maximum possible magnification across all FFP planet masses is $\lesssim$100 and never exceeds 2 for Earth-mass planets.
This will make it difficult to securely detect the event from two observatories and to also robustly measure {\piE}.
The limiting factor shifts when considering main sequence sources, with ${\rs} \lesssim 1{\rsun}$.
A smaller $\rho$ indicates a higher possible magnification but a correspondingly lower probability for measuring finite-source effects.
Given a Solar-type source with ${\rs} = 1{\rsun}$ and an FFP lens at {\dl} = 4 kpc, there is only a $\sim$30$\%$ chance of measuring $\rho$ for a planet with ${\mffp} = 1 {\rm M}_{\oplus}$, and this drops to $\sim$1$\%$ for ${\mffp} = 1000 {\rm M}_{\oplus}$.

\vspace{10.0mm}
\section{Constraining Lens Flux {\fl} to Exclude Possible Host Star} \label{sec:fl}

Lastly, once a planetary-mass lens has been identified, it is necessary to exclude the possibility of the FFP being bound by searching for and constraining the flux contribution from a possible host star.
This requires a measurement of the source flux and at least one epoch of high-resolution data.
The assumed procedure is as follows.

\subsection{General Methodology} \label{sec:fl_method}

The flux $F$ of a microlensing event at time $t$ is given by:
\begin{equation} \label{eq:flux_ulens}
   F(t) = {\fsource}A(t) + {\fb},
\end{equation}
where {\fsource} is the flux of the source, $A(t)$ is the magnification of the source at $t$, and {\fb} is the blend flux of all other stars that are not resolved.
It is relatively straightforward to fit a microlensing model for a well-sampled light curve of a single-lens microlensing event, giving $A(t)$ for all times $t$.
This is true regardless of observing bandpass, since microlensing is achromatic.
Then, for every additional band beyond the primary one with which $A(t)$ is computed, in principle only two data points, taken at two different magnifications, are required to measure {\fsource} and {\fb} in that band.

A high-resolution image of the microlensing event taken at any time will resolve out all stars not dynamically associated with the event to a high probability, meaning that {\fb} will consist only of the lens and companions to the lens and/or source.
Taking a high-resolution observation at baseline, when the event is over and the source is no longer magnified, optimizes the chances of detecting light from any luminous lenses, since the flux contribution from the source will be minimized and no ambient interloping stars will be contributing to {\fb}.
We will refer to such a high-resolution flux measurement as {\ftbase}, since only light from the microlensing target --- the lens and source systems --- will be included in the point-spread function (PSF).
This can be combined with the known source flux to determine {\deltaF}, a measure of any flux in excess of the known source flux in the high-resolution image, via:
\begin{equation} \label{eq:deltaF}
    {\deltaF} \equiv {\ftbase} - {\fsource}.
\end{equation}
Regarding FFPs, {\deltaF} = 0 provides a limit on the brightness of a possible lens host star.
Since the distances to the FFPs will be derived from the measurements of {\piE} and {\thetaE}, this allows for the strongest constraints on the lens host star flux {\fl}, which in turn provides yields a limit on the lens host star mass {\ml}.

This same principle has been used to characterize a handful of bound planetary microlensing events  \citep{bennett2007,dong2009b,janczak2010,sumi2010,batista2011,batista2014,batista2015,fukui2015}.
The high-resolution data are typically taken in $H$-band, which balances the seeing-limited angular resolution with the attainable Strehl ratio and the sky background.
Hereafter we will consider only $H$-band fluxes of the source, lens, and target, {\hs}, {\hl}, and {\htar}, respectively, but note that this technique works for any observational band.

Estimating the mass above which it is possible to rule out the presence of a lens host star requires a careful treatment of the relevant uncertainties.
Here we consider two observational regimes.

\subsection{Fiducial Simulation} \label{sec:fiducial}

\begin{figure*}
   \centerline{
      \includegraphics[width=18cm]{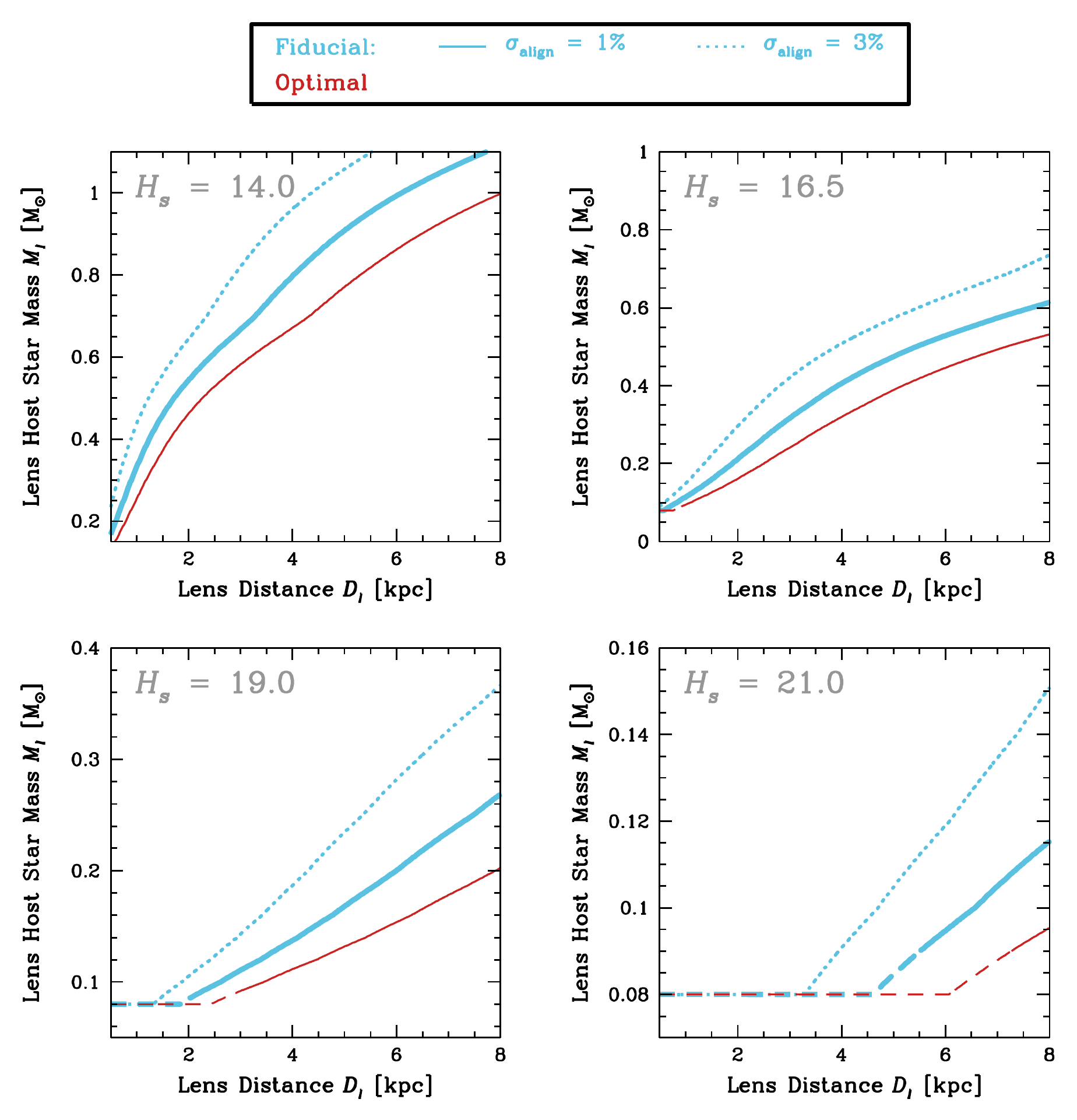}
   }
   \caption{
      \footnotesize{
         The highest lens mass {\ml} of a possible host star that can be excluded, as a function of lens distance {\dl}.
         Each panel represents a source star with a different apparent $H$-band brightness, corresponding to the four broad categories of spectral types discussed in \S \ref{sec:rho}.
         For each, the blue line corresponds to our fiducial simulation (see \S \ref{sec:fiducial}) assuming a conservative 3$\%$ precision in photometrically aligning the high-resolution data (dotted line) and a 1$\%$ precision (solid line).
         The red line shows the results for our optimal simulation (see \S \ref{sec:optimal}).
         We use a 5 Gyr Padova isochrone \citep{bressan2012,chen2014,tang2014,chen2015} that extends only to 0.09${\rm M}_{\odot}$, so we extrapolate it to the deuterium-burning limit (dashed lines), at 0.08${\rm M}_{\odot}$, which we set as the limit for stellar-mass lens host star exclusion.
         If ${\hs} \gtrsim 19.0$ it is possible to completely exclude the presence of a stellar host out to a certain lens distance threshold.
         For ${\hs} \approx 19$ using our fiducial simulation and assuming a photometric alignment precision of 1$\%$, which is a good representation of the observing capabilities for {\ktcn}, we find that stellar lenses can be excluded up to ${\dl} \approx 2$ kpc.
         Our optimal simulation extends this threshold to $\sim$6 kpc for ${\hs} \approx 21$, which is typical for {\wfirst} sources.
         \vspace{1.0mm}
      }
   }
   \label{fig:fl}
\end{figure*}

For our fiducial simulation, we assume the observational resources that will be implemented during {\ktcn}.
A proposal was accepted to take target-of-opportunity (ToO) observations of short-timescale FFP candidates using NIRC2 on Keck 2.
Each ToO will be used to obtain a high-resolution measurement of the $H$-band flux of a microlensing target while the source is magnified, which we will refer to as {\htmag}.
We also assume the existence of a future high-resolution epoch of the target when the source is at its baseline flux, which we will denote as {\htbase}.
As described in \S \ref{sec:fl_method}, these two points are sufficient to measure the source flux ({\hshighres}) and blend flux ({\hbhighres}) in the high-resolution data, given a light curve model derived using other bandpasses.
Equation (\ref{eq:deltaF}) refers to the general case for which only the baseline epoch of high-resolution data exists.
In this scenario, however, obtaining a magnified point \textit{and} an eventual baseline epoch mean that a measurement of $H_{b, \rm{high-res}} = 0$ establishes a limit on the brightness of a possible lens host star.
Under this procedure, the associated uncertainties are thus:
\begin{enumerate}
   \item $\sigma_{\htmag}$: the statistical uncertainty of the microlensing target taken using the high-resolution facility when the source is magnified;
   \item $\sigma_{\htbase}$: the statistical uncertainty of the microlensing target taken using the high-resolution facility when the event is over and the source has returned to its baseline brightness;
   \item $\sigma_{\rm align}$: the uncertainty of photometrically aligning the two high-resolution images in order to compute {\hbhighres};
   \item $\sigma_{\hbhighres}$: the statistical uncertainty of the blend brightness, which includes its covariance with the other parameters from the light curve model; and
   \item $\sigma_{\rm calib}$: the uncertainty required to calibrate to a photometric standard system.
\end{enumerate}

For our fiducial computation, we assume $\sigma_{\htmag} = \sigma_{\htbase} = 0.1\%$ and $\sigma_{\rm align} = \sigma_{\hbhighres} = 1\%$.
We assume that $\sigma_{\rm calib} = 1\%$ \citep{batista2014} and consider four different $H$-band source magnitudes, each of which corresponds to one of the four spectral types explored in \S \ref{sec:rho}.
Assuming a source distance of {\ds} = 8.2 kpc \citep{nataf2013} and an $H$-band extinction of $A_{H} = 0.5$, we take {\hs} = 14.0 for a giant source, {\hs} = 16.5 for a subgiant, {\hs} = 19.0 for a Solar-type main sequence source, and {\hs} = 21.0 for a late K/early M dwarf.
We add all uncertainties in quadrature and determine the $H$-band apparent magnitude of a possible lens host star, {\hl}, that can be excluded at the three-sigma level, from an upper limit of {\hbhighres} = 0.
Finally, using a 5 Gyr Padova isochrone \citep{bressan2012,chen2014,tang2014,chen2015}, we convert {\hl} into a limit on the lens host star mass, {\ml}, as a function of lens distance {\dl}.
The isochrone extends only to 0.09${\rm M}_{\odot}$, so we extrapolate down to the deuterium-burning limit at 0.08${\rm M}_{\odot}$, which we establish as the threshold for stellar-mass FFP host star exclusion.
Our results are shown in Figure \ref{fig:fl}.

Giant and subgiant sources are generally too bright to be able to robustly rule out the presence of an M-dwarf FFP host star, particularly at distances of several kiloparsecs.
However, for fainter main sequence sources, for which the flux contribution from a potential lens host star will be a significantly larger fraction of the total light measured in the baseline high-resolution observation {\ftbase}, the situation is much different.
Assuming ${\hs} \approx 19$, stellar hosts of any mass can be excluded up to $\sim$1.8 kpc, and even for FFPs in the bulge the most massive host star that could go undetected is $\sim$0.25 {\msun}.
Finally, as a more conservative limit, we set $\sigma_{\rm align} = 3\%$, the results for which are also shown in Figure \ref{fig:fl}.
This decreases the limiting distance closer than which hosts of any mass can be excluded to $\sim$1.3 kpc.
Correspondingly, the mass limit that can be ruled out for an FFP lens at any distance increases to $\sim$0.37 {\msun}.

\subsection{Optimal Simulation} \label{sec:optimal}

We also perform a second simulation with an optimal set of assumptions regarding the relevant uncertainties.
The primary band in which {\wfirst} will conduct its microlensing survey, W149, will have a PSF of $\sim$0.14$\arcsec$ \citep{spergel2015}.
At this angular resolution, the probability of having an ambient interloping star with $H_{\rm AB} < 26$ blended with the microlensing event is $<$15$\%$ \citep{spergel2015}.
Thus, for the vast majority of microlensing events that {\wfirst} will detect, having additional high-resolution observations taken with a different facility, both while the source is magnified and while it is at baseline, is unnecessary.
Moreover, during a microlensing season images will be taken in W149 every $\sim$15 minutes, producing light curves with exquisite photometric precision and sampling.
Lastly, {\wfirst}'s microlensing predicted detection rates peak between ${\rm W149}_{AB}$ = 22--23 (Penny et al., in prep.), much fainter than for {\ktcn} or current ground-based surveys.
In summary, the {\wfirst} data set will largely be self-contained.
Given these factors, for our optimistic simulation we remove $\sigma_{\rm align}$ and $\sigma_{\htmag}$, set ${\hbhighres} = \sigma_{\htbase} = 0.1\%$, and assume $\sigma_{\rm calib} = 1\%$.

Figure \ref{fig:fl} shows the results for our optimal scenario.
For subgiant and giant sources it is still difficult to set meaningful mass limits for a lens host star, particularly as {\dl} increases.
However, for Solar-type main sequence sources with ${\hs} = 19.0$, the distance to which stellar hosts can be excluded increases to $\sim$2.4 kpc, and for lenses in the bulge we can probe down to $\sim$0.2{\msun}.
This improves dramatically for late K/early M dwarf sources with ${\hs} = 21.0$.
Stellar lens host stars can be excluded out to $\sim$6.1 kpc, and even in the central bulge the highest mass star that would remain undetected is only $\sim$0.1{\msun}.

\subsection{Spatially Resolve Lens and Source} \label{sec:resolve}

For our simulations and results regarding lens flux constraints we have operated in the regime of prompt follow-up photometry, i.e., high-resolution images taken shortly after the event is over.
Even with facilities such as NIRC2 on Keck and {\wfirst}, the lens and source will remain unresolved.
However, it is possible to set even stronger {\fl} constraints by waiting for the lens and source to separate according to their relative proper motion {\murel}.
Then the possible lens host star can be treated as an isolated object, allowing for more robust {\fl} and, ultimately {\ml} constraints when compared to searching for an increase in flux on top of the (known) source flux.
\citet{gould2016} finds that for FFP candidate events it is feasible to exclude the presence of lens host stars with separations comparable to the Oort Cloud using existing adaptive optics facilities by waiting $\sim$50 years after the event is over.

This technique still requires that {\piE} and {\thetaE} (from a determination of $\rho$) are measured, such that the mass of the FFP will be known.
By coupling these parameters with {\tE}, the lens-source relative proper motion vector {\murelvec} can be measured.
It is then straightforward to estimate the timescale on which the lens and source could be spatially resolved using a given facility as well as the direction of separation (see \citealt{drsexypants:2015a} for a more thorough discussion of spatially resolving lenses and sources).
It would still be optimal to obtain a prompt high-resolution epoch shortly after the event is over.
Given the stellar surface density toward the bulge, it is best to identify and exclude any ambient interloping stars that could possible be confused for the microlensing target, especially if the proper motion is not well constrained.

\section{Discussion} \label{sec:discussion}

Here we present the implications our findings have for the characterization of FFP candidate events.
It is eminently possible to explore the frequency, Galactic distribution, and even mass function of FFPs without robustly measuring {\piE} and {\thetaE} (via $\rho$), which together provide mass and distance measurements for individual events.
Nor is fully excluding stellar-mass hosts a requirement for gaining traction in understanding FFP demographics.
In this section, however, we will focus on the full set of constraints as they can be measured using data taken during {\ktcn} or by {\wfirst}.

\begin{figure*}
   \centerline{
      \includegraphics[width=18cm]{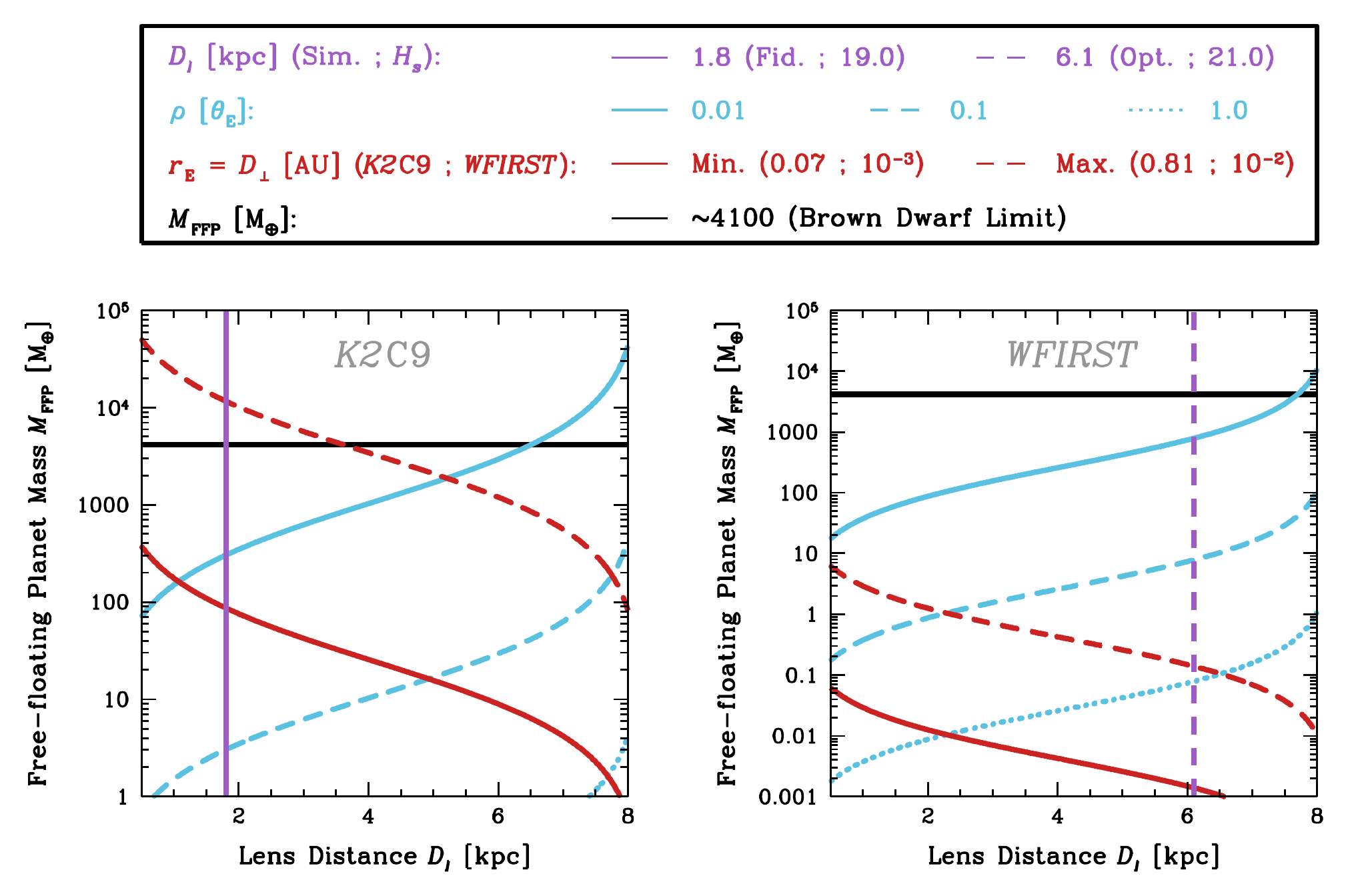}
   }
   \caption{
      \footnotesize{
         Free-floating planet mass {\mffp} as a function of lens distance {\dl} for {\ktcn} (left panel) and {\wfirst} (right panel).
         The red lines correspond to the minimum and maximum values of the projected separation {\dproj} for the respective satellites throughout their microlensing campaigns.
         The blue lines indicate $\rho$ = 0.01 (solid lines), 0.1 (dashed), and 1.0 (dotted).
         In the right panel they correspond to a physical source radius of ${\rs} = 0.5{\rsun}$, typical for {\wfirst} sources.
         In the left panel we instead assume ${\rs} = 1.0{\rsun}$, since ${\hs} \approx 21$ will likely be too faint for {\ktcn} but ${\hs} \approx 19$ allows for the otherwise strongest lens flux constraints.
         The purple line indicates the largest lens distance {\dl} at which all stellar-mass lens host stars can be excluded for our fiducial simulation assuming an $H$-band source magnitude of {\hs} = 19 (solid; left panel), and for our optimal simulation assuming {\hs} = 21 (dashed; right panel).
         Finally, the black horizontal line demarcates the deuterium-burning limit that is the classical boundary between planets and brown dwarfs.
         Both panels show how the constraints for measuring {\piE} (set by the red lines) and $\rho$ (blue lines) work in opposing directions.
         Regarding {\ktcn}, FFP events occurring toward the beginning of the campaign have a higher probability of being fully characterized.
         In the case of {\wfirst}, on the other hand, full charactaerization is possible for FFPs with masses at least as low as super-Earths throughout its entire microlensing observing window.
         \vspace{1.0mm}
      }
   }
   \label{fig:mffp_dl}
\end{figure*}

\subsection{Application to {\ktcn}} \label{sec:applications_k2c9}

The projected separation {\dproj} of {\ktcn} reaches its minimum value of 0.07 AU one week into the campaign, on 14/April/2016, after which it increases monotonically to 0.81 AU on 1/July.
This significantly affects its ability to measure {\piE} for FFP lenses with different masses and distances throughout the campaign.
Figure \ref{fig:mffp_dl} shows {\mffp} as a function of {\dl}.
By assuming a value of {\rEproj} that is equal to {\dproj} it is possible to determine the masses and distances of FFPs that will yield detectable events at a given time during {\ktcn}.
As discussed in \S \ref{sec:rEproj}, the probability that events with ${\dproj} = {\rEproj}$ will be detectable by the pair of observer locations is $\sim$60$\%$, and this probability declines steeply as ${\dproj}/{\rEproj}$ increases.
We thus use this as a benchmark for sensitivity to {\piE}.
Immediately we see that toward the end of {\ktcn} {\piE} can only be measured for lenses with mass lower than the deuterium-burning limit for ${\dl} \gtrsim 4$ kpc.
Near the beginning of the campaign it will be possible to measure a parallactic shift for nearby Jupiter-mass lenses, with its sensitivity extending down to Earth-mass FFPs as the lens distance increases to ${\dl} \approx 8$ kpc.

But {\piE} is only one of two ingredients necessary for measuring {\mffp} and {\dl}.
The other is $\rho$, which helps determine {\thetaE}.
By rearranging the right-hand side of Equation (\ref{eq:rho_prob}) we compute {\mffp} as a function of {\dl} for different values of $\rho$, given an assumed physical source radius {\rs} (and thus angular source radius {\thetastar}, assuming ${\ds} = 8.2$ kpc).
To compute $\rho$ for {\ktcn} (left panel of Figure \ref{fig:mffp_dl}) we assume ${\rs} = 1.0{\rsun}$.
We select this as the benchmark value of {\rs} because ${\hs} \approx 21$, corresponding to ${\rs} = 0.5{\rsun}$, is likely too faint for {\ktcn} and ${\hs} \approx 19$ allows for the otherwise strongest lens flux constraints.
From the left-hand side of Equation (\ref{eq:rho_prob}), the probability of detecting finite-source effects, and thus measuring $\rho$ and, ultimately, determining {\thetaE}, scales approximately as $\rho$ itself.
Thus, the value of $\rho$ at a given ({\dl}, {\mffp}) coordinate gives the probability of measuring $\rho$ for an FFP with those physical properties.
Rephrased, $\rho^{-1}$ indicates, roughly, the number of events with that specific ({\dl}, {\mffp}) combination that would need to be detected in order to measure $\rho$ for at least one of them.
The constraints for measuring {\piE} and $\rho$ work in opposing directions.
For a fixed {\dproj}, the mass to which the satellite is sensitive decreases as {\dl} increases.
However, as {\dl} increases, the probability of measuring $\rho$ \textit{increases}, for a fixed FFP mass.

Finally, we incorporate the flux constraint for ruling out the presence of lens host stars.
From the left panel of Figure \ref{fig:mffp_dl}, the largest {\dl} at which stellar-mass lens host stars can be excluded for sources with ${\hs} \approx 19$ is $\sim$1.8 kpc from our fiducial simulation.
When considering all three factors, {\piE}, $\rho$, and {\fl}, the optimal regime for characterizing FFPs occurs for the smallest {\dproj}, when the sensitivity extends to FFPs with the lowest masses, for a fixed {\dl}, and for when the probability of measuring $\rho$ is the highest, i.e., larger values of $\rho$.
This drives the sensitivity of {\ktcn} toward the first few weeks of the campaign, when ${\dproj} \lesssim 0.2$ AU, as is furthermore supported by the goal of excluding stellar-mass host stars, which requires that ${\dl} \lesssim 1.8$ kpc.
In this regime, however, while stellar hosts can be excluded at the three-sigma level from {\fl} constraints, the probability of measuring both {\piE} and $\rho$ is only $\sim$1.5$\%$.
At the end of {\ktcn}, when ${\dproj} = 0.81$ AU, the lowest mass lens for which it will be possible to measure {\piE} and also constrain the flux will be $M \approx 10^{4}{\rm M}_{\oplus}$, above the brown dwarf mass limit.

\subsection{Planning for {\wfirst}} \label{sec:applications_wfirst}

The current plan is for each of {\wfirst}'s 72-day observing seasons to be centered on an equinox.
Its projected separation would thus be restricted to $0.006 \lesssim {\dproj}/{\rm AU} \lesssim 0.009$.
Nevertheless, given the uncertain nature of the orbital parameters (see \S \ref{sec:dproj}) and the exact timing of the observing seasons, we presume $D_{\perp, {\rm min}} \approx 10^{-2}$ AU and $D_{\perp, {\rm max}} \approx 10^{-3}$ AU to bracket the full range possible for {\wfirst}.
These are shown in the right panel of Figure \ref{fig:mffp_dl}.
From the detectability limit of ${\dproj} = {\rEproj}$ discussed in \S \ref{sec:rEproj}, this indicates that {\wfirst} will be able to measure {\piE} for Earth-mass planets for ${\dl} \gtrsim 2$ kpc and for lower-mass planets at larger distances.
Were the Sun-angle restrictions relaxed such that {\wfirst} could observe toward the bulge near the June solstice, when its projected separation could reach as small as 10$^{-3}$ AU, it would be sensitive to planets $\sim$2 orders-of-magnitude less massive (i.e., free-floating moons).

The typical source stars for {\wfirst} will have ${\rs} = 0.5{\rsun}$, meaning that the probability to measure $\rho$ is $\gtrsim$10$\%$ for Earth-mass FFPs with ${\dl} \gtrsim 2.5$ kpc.
The corresponding source magnitude of ${\hs} \approx 21$, combined with the optimal lens flux simulation (which is appropriate for {\wfirst}), suggests that it will be possible to exclude host stars for FFPs detected by {\wfirst} for ${\dl}\lesssim 6.1$ kpc.
This means that {\wfirst} will be sensitive to FFPs at least down to the mass of super-Earths throughout the Galaxy.

One possible limiting factor, however, could be the capabilities of the accompanying ground-based facilities.
A source with ${\hs} \approx 21$ will have an optical magnitude of $I \approx$ 22--24.
This is below the sensitivity of current microlensing surveys.
Additionally, the maximum magnification limitation from having a larger $\rho$ means that many of these events may not get sufficiently bright to be detected even at peak.
However, if future surveys will be conducted by larger telescopes, such as that proposed for Subaru, this limit is no longer relevant.
Secondly, the bulge is visible only for $\sim$1--7 hours from CTIO (as a representative for the southern hemisphere) during the currently planned {\wfirst} campaigns (as shown on the bottom panel of Figure \ref{fig:dproj}).
For telescopes in the northern hemisphere, such as Subaru on Mauna Kea, the bulge is visible for only 60$\%$ of each {\wfirst} observing season and even then only up to a maximum of $\sim$3 hours per night.

\acknowledgments

Work by CBH and YS was supported by an appointment to the NASA Postdoctoral Program at the Jet Propulsion Laboratory, administered by Universities Space Research Association through a contract with NASA.
We thank David Ciardi, Chas Beichman, and Chris Gelino for useful discussions regarding adaptive optics observations.
Finally, we thank Scott Gaudi, Dan Maoz, Matt Penny, and Sebastiano Calchi Novati for a careful and fruitful reading of the manuscript.


\bibliographystyle{apj}

\end{document}